\documentclass[12pt,reqno]{amsart}
\usepackage{amsmath}
\usepackage{amssymb}

\numberwithin{equation}{section}
\newtheorem{rmk}{Remark}[section]

\newcommand{\barh}{{\bar h}}
\newcommand{\g}{{{\bf g}}}

\let\pa=\partial

\begin{document}

\title[Noncommutative gravitational collapse]
{Gravitational collapse of spherically symmetric stars
in \\ noncommutative general relativity}

\author{Wen Sun, Ding Wang, Naqing Xie,
R. B. Zhang and Xiao Zhang}

\address[W. Sun]{Lianyungang Teachers College,
Jiangsu 222006, China}

\address[D. Wang, X. Zhang]{Institute of Mathematics,
Academy of Mathematics and Systems Science, Chinese Academy of
Sciences, Beijing 100190, China}
\email{wangding@amss.ac.cn, xzhang@amss.ac.cn}

\address[N. Xie]{Institute of Mathematics,
School of Mathematical Sciences, Fudan
University, Shanghai 200433, China}
\email{nqxie@fudan.edu.cn}

\address[R. B. Zhang]{School of Mathematics
and Statistics, University of Sydney, Sydney,
NSW 2006, Australia}
\email{rzhang@maths.usyd.edu.au}

\date{}

\begin{abstract}
Gravitational collapse of a class of spherically
symmetric stars are investigated.  We quantise the
geometries describing the gravitational collapse by a
deformation quantisation procedure.
This gives rise to noncommutative spacetimes with
gravitational collapse.

\bigskip

\noindent{\bf PACS.}
04.70.Bw\hspace{1mm} Classical black holes -
04.70.Dy\hspace{1mm} Quantum aspects of black holes -
11.10.Nx  Noncommutative field theory.
\end{abstract}

\maketitle

\section{Introduction}\label{Intro}

Gravitational collapse \cite{ABS}-\cite{W} is one of the most
dramatic phenomena in the universe. When the pressure is not
sufficient to balance the gravitational attraction inside a star,
the star undergoes sudden gravitational collapse possibly
accompanied by a supernova explosion, reducing to a super dense
object such as a neutron star or black hole.

Theoretical investigations predicting gravitational collapse were
carried out in the early 1930s in the ground breaking work of
Chandrasekhar \cite{C1, C2}.  In 1939, Oppenheimer and Snyder
\cite{OS} investigated the collapse process of ideal spherically
symmetric stars equipped with the Tolman metric \cite{T} $-d\tau ^2
+e ^{\bar \omega(\tau, R)} dR^2 +e ^{\omega(\tau, R)}\big(d\theta
^2+\sin^2 \theta d\psi^2\big)$. When the energy-momentum of an ideal
star is assumed to be given by perfect fluids, Tolman's metric
allows the case of dust which has zero pressure. In the dust case,
Oppenheimer and Snyder solved the Einstein field equations by
further assuming that the energy density is constant. They showed
that stars above the Tolman-Oppenheimer-Volkoff mass limit \cite{OS,
OV} (approximately three solar masses) would collapse into black
holes for reasons given by Chandrasekhar. The work of Oppenheimer
and Snyder also marked the beginning of the modern theory of black
holes.

The purpose of the present paper is to investigate gravitational
collapse in noncommutative general relativity. We shall work within
the framework of the noncommutative Moyal geometry developed in
\cite{CTZZ, WZZ1, WZZ2}. Gravitational collapse is in principle
understood classically, however, quantum effects will become
important at the final stage of the collapse, especially if a star
collapses into a black hole. As there still lacks a consist theory
of quantum gravity, it is useful to incorporate some quantum effects
into gravity by deforming general relativity. Much effort has been
made in this area in recent years, resulting in several tentative
proposals for noncommutative general relativity
\cite{Cham1}-\cite{Wess2}, \cite{CTZZ, WZZ1, WZZ2},
\cite{Madore1}-\cite{Ruiz2}.
Much work has been done to investigate noncommutative corrections to
black holes, and we refer to\cite{WZZ1}, \cite{CT1}-\cite{Random5} for details.

On a cautionary note, we should mention that the widely cited
papers \cite{Wess1, Wess2} were shown in \cite{AMV} to yield results
entirely different from the low energy limit of string theory.
If the cause of this is not mathematical imprecisions in \cite{Wess1, Wess2},
then it may be an indication that there are flaws in the rationale of these papers,
as the low energy limit of string theory for an appropriate
choice of vacuum is physically realistic.

In the present paper, we first investigate static interior solutions of a class of
spherically symmetric stars equipped with the Misner-Zapolsky metric
\cite{MZ} (see equation \eqref{XZ}) and have energy-momentum given
by perfect fluids (see equation \eqref{fluid}). The energy densities
of the stars are taken to be decreasing functions instead of
(physically unrealistic) constant functions.  New solutions of the
Einstein field equations are obtained, and their singularities
signalling gravitational collapse are discussed.
We feel that these results are interesting by themselves
even from the point of view of classical gravity.

We then quantise the metrics obtained by the deformation
quantisation procedure developed in \cite{CTZZ, WZZ1, WZZ2}. This
gives rise to noncommutative spacetimes with static interior singularity.

Finally, we quantise the dust solutions studied by Oppenheimer and Snyder
\cite{OS}.  It gives rise to the noncommutative dynamical gravitational
collapse. As far as we are aware, gravitational collapse in the
noncommutative setting has not been studied before.

\section{Spherically symmetric stars} \label{Spher}

Suppose a spherically symmetric star is equipped with the following
metric
\begin{eqnarray}
g=-e^{2\alpha(t,r)}dt^2+e^{2\beta(t,r)}dr^2+r^2(d\theta^2+\sin^2\theta
d\psi^2),  \label{XZ}
\end{eqnarray}
and its energy-momentum is given by perfect fluids with
\begin{eqnarray}\label{fluid}
T_{\mu\nu}=(\varepsilon +p )U_\mu U_\nu +pg_{\mu\nu},
\end{eqnarray}
where $\varepsilon$ is the energy density, $p$ the pressure and the
vector $U$ is the 4-velocity of the fluid elements. This metric was
studied by Misner and Zapolsky \cite{MZ} for neutron star models.
Its physical properties differ considerably from those of the Tolman
metric \cite{T}. In particular, it rules out dust solutions (with
$p=0$ but $\varepsilon\ne 0$). We shall show that, for general spherical symmetric
metric (\ref{XZ}), if the energy density is independent of the time
$t$, then the metric must be static. We shall also study the cases
that energy densities are given by step functions and quadratic
decay functions.

We mention that in some textbooks, e.g. \cite{HE, KSHM, W},
the metric (\ref{XZ}) was discussed in the static case with functions
$\alpha$, $\beta$ depending on $r$ only. The appearance of a curvature
singularity was interpreted as an indication of gravitational collapse.

We choose the co-moving coordinates and take the four-velocity $U$
to be pointing in the timelike direction, that is, $U_1 =U _2 =U _3
=0$. We also normalise $U$ by setting $U^\nu U _\nu =-1$. Now the
components of energy-momentum tensor in the coordinates
$(t,r,\theta,\psi)$ are given by
 \begin{eqnarray*}
T _{\mu \nu}=\left(\begin{array}{cccc}
  e^{2\alpha}\varepsilon &  &  &  \\
   & e^{2\beta}p &  &  \\
   &  & r^2p &  \\
   &  &  & r^2 \sin^2\theta p
\end{array}\right).
 \end{eqnarray*}
Denote the Einstein tensor by
$G_{\mu\nu}={R}_{\mu\nu}-\frac{R}{2}g_{\mu\nu}$. The Einstein field
equations give that \cite{S}
 \begin{eqnarray*}
 \begin{aligned}
    G_t ^t=&e^{-2 \beta}\Big(\frac{1}{r^2}-\frac{2\pa _r \beta}{r}\Big)-\frac{1}{r^2}
          =-\varepsilon,\\
    G_r ^r=&e^{-2 \beta}\Big(\frac{1}{r^2}+\frac{2\pa _r \alpha}{r}\Big)-\frac{1}{r^2}
          =p,\\
    G_t ^r=&e^{-2 \beta}\frac{2\pa _t \beta}{r}=0,\\
    G _\theta ^\theta =&G_\psi ^\psi =e^{-2 \beta}\Big(\pa _r ^2 \alpha +(\pa _r \alpha) ^2
                        +\frac{1}{r}(\pa _r \alpha -\pa _r \beta)-\pa _r \alpha \pa _r \beta\Big)\\
                       &-e^{-2\alpha}\Big(\pa _t ^2 \beta
                       +(\pa _t \beta) ^2 -\pa _t \alpha \pa _t \beta\Big)=p.
 \end{aligned}
 \end{eqnarray*}

As a consequence of the Bianchi identity, we have the
Tolman-Oppenheimer-Volkoff (TOV) equation \cite{ABS,OV,T}
\begin{eqnarray} \label{TOV} (\varepsilon+p)\pa _r \alpha + \pa _r p =0. \end{eqnarray}

We now show that the metric (\ref{XZ}) does not allow any dust
solution: the vanishing of the pressure $p$ implies the vanishing of
the energy density $\varepsilon$. In fact, (\ref{TOV}) implies
either $\varepsilon =0$ or $\pa _r \alpha =0$ if $p=0$. In the
latter case the $G_r ^r$ equation gives $\beta =0$, thus the $G_t
^t$ equation gives $\varepsilon =0$. By the Birkhoff theorem,
(\ref{XZ}) must be Schwarzschild in this case. This is different
from the Tolman metric, for which the dust solution was constructed
by Oppenheimer and Snyder \cite{OS}.

It is straightforward that the $G _\theta ^\theta$ and $G_\psi
^\psi$ equations are consequence of those of $G_t ^t$, $G_r ^r$,
$G_t ^r$ and the TOV equation (\ref{TOV}) (see, e.g,
\cite{HE,KSHM,W} for the proof of this fact in the static case).  In
fact, the $G_t ^t$, $G_r ^r$ equations give
 \begin{eqnarray*}
 \begin{aligned}
\partial_r \beta&=\frac{1}{2r}+\big(\frac{\varepsilon
r}{2}-\frac{1}{2r}\big)e^{2\beta},\\
\partial_r \alpha&=-\frac{1}{2r} +\big(\frac{pr}{2}+\frac{1}{2r}\big)e^{2\beta}.
 \end{aligned}
 \end{eqnarray*}
From the $G_t ^r$ equation, we obtain
 \begin{eqnarray*}
\beta =\beta(r).
 \end{eqnarray*}
This implies that
 \begin{eqnarray}
\varepsilon=\varepsilon(r). \label{varp-r}
 \end{eqnarray}
From the TOV equation (\ref{TOV}), we obtain
 \begin{eqnarray*}
 \begin{aligned}
\partial^2_r\alpha=& e^{4\beta} \big(\frac{\varepsilon p
r^2}{4}+\frac{\varepsilon}{4}-\frac{3p}{4}-\frac{1}{2r^2}-\frac{r^2p^2}{4}\big)\\
&+e^{2\beta}\big(\frac{5p}{4}+\frac{\varepsilon}{4}\big)+\frac{1}{2r^2}.
 \end{aligned}
 \end{eqnarray*}
Substituting these into the $G_\theta ^\theta$ equation, we find that the
left hand side is equal to $p$, so it is an identity.

Now we show that if the state equation $p=p(\varepsilon)$ holds, the
metric (\ref{XZ}) must be static. In fact, from (\ref{varp-r}), we
know that $p=p(r)$. Using the $G_{r}^r$ equation, we obtain
 \begin{eqnarray*}
\pa_t \pa _r \alpha =0.
 \end{eqnarray*}
This implies
 \begin{eqnarray*}
\alpha =f(r)+h(t).
 \end{eqnarray*}
Replacing $t$ by $\bar t=\int e^h dt$, the metric (\ref{XZ}) can be
rewritten as
 \begin{eqnarray}
{g}=-e^{2f(r)}d\bar t
^2+e^{2\beta(r)}dr^2+r^2(d\theta^2+\sin^2\theta d\psi^2).
\label{XZ-1}
 \end{eqnarray}
Since $ G_{tt}=G_{\bar t \bar t} \left(\frac{d\bar t}{dt} \right)^2
=e^{2\alpha}\varepsilon$ if and only if $G_{\bar t \bar
t}=e^{2f}\varepsilon, $ we conclude that the fields are static in
the sense that they do not depend on $\bar t$. Thus we can assume
$\alpha =\alpha (r)$ by replacing $f$ by $\alpha$ in (\ref{XZ-1}).

A spherically symmetric star may be assumed to be a ball of
radius $r_0$ centered at $r=0$. Then the energy density
$\varepsilon$ and the pressure $p$ must satisfy
 \begin{eqnarray*}
 \begin{aligned}
\varepsilon &=0 \quad \mbox{if} \quad r > r_0,\\
 p&=0 \quad \mbox{if} \quad r \geq r_0.
 \end{aligned}
 \end{eqnarray*}

From now on we assume that the pressure of the star depends on $r$
only. Thus the star is static by the above discussion. Let
$e^{-2\beta(r)}=1-\frac{2m(r)}{r}$ for some $m(r)$, then \begin{eqnarray}
m(r)=\frac{1}{2}(r-re^{-2\beta}). \label{beta} \end{eqnarray}  The $G_{r} ^r$
equation gives
 \begin{eqnarray}
\frac{d \alpha}{dr}=\frac{r^3p+2m(r)}{2r(r-2m(r))}, \label{alpha}
 \end{eqnarray}
and the $G_{t} ^t$ equation leads to
 \begin{eqnarray*}
\frac{d m}{dr}=\frac{1}{2}r^2 \varepsilon.
 \end{eqnarray*}
If $m(0)\neq 0$, then $e^{2\beta (0)} =0$ by (\ref{beta}) and the
metric $g$ degenerates at $r=0$. Exclude this case, we may assume
 \begin{eqnarray}
m(0)=0. \label{m0}
 \end{eqnarray}
Finally, in terms of the contracted Einstein equation, we have
 \begin{eqnarray} \label{cein}
R(g)=\varepsilon(r)-3p(r).
 \end{eqnarray}

\section{Interior solutions} \label{Decreasing}

In the original investigation on
spherically symmetric stars equipped with the Tolman metric in
\cite{OS}, the energy density was assumed to be a constant.
In view of the fact  that the energy density should decrease as
$r$ becomes large, Gu \cite{G} studied the spherically symmetric dust
with the energy density given by a (decreasing) step function. In this
case he found that singularity could only appear away from the origin.
The complete classification of the spherically symmetric dust was given
by Hu \cite{H}. In this section, we analyze the results of Section \ref{Spher} in
detail for two classes of spherically symmetric stars equipped with
the Misner-Zapolsky metric \cite{MZ} and has decreasing energy
density functions.

\subsection{Constant and step function energy densities}

We first assume that
\begin{eqnarray}
\varepsilon(r)=\varepsilon=\mbox{constant}>0 \quad \mbox{for}
\quad r \leq r_0,
\end{eqnarray}
and the mass function is
\begin{eqnarray}
m(r)=\Big\{\begin{array}{ll}
\frac{\varepsilon}{6}r^3 & \ \mbox{for} \quad r \leq r_0\\
\frac{\varepsilon}{6}r_0 ^3 & \ \mbox{for} \quad r > r_0.
\end{array}
\end{eqnarray}
Then the TOV equation (\ref{TOV}) with boundary condition
$p(r_0)=0$ gives the following explicit formula for the pressure
$p(r)$ inside the star \begin{eqnarray} \label{pre} p(r)=\varepsilon
\frac{\sqrt{3-\varepsilon r^2} -\sqrt{3-\varepsilon r_0
^2}}{3\sqrt{3-\varepsilon r_0 ^2} -\sqrt{3-\varepsilon r^2}} \quad
\mbox{for} \quad r \leq r_0. \end{eqnarray}  By Birkhoff theorem \cite{ABS},
the exterior metric must be Schwarzschild solution with mass
$\frac{\varepsilon r_0 ^3}{6}$: \begin{eqnarray} \label{ext}
\begin{aligned}
g_{\mbox{ext}}=&-\left(1-\frac{\varepsilon r_0 ^3}{3r}\right)dt ^2 +
\left(1-\frac{\varepsilon r_0 ^3}{3r}\right)^{-1} dr ^2\\
           &+r ^2 \left(d \theta ^2 +\sin ^2 \theta d \psi ^2\right)
\end{aligned}
\end{eqnarray}
for $r > r_0$. Solving (\ref{alpha}) with the continuity condition
\begin{eqnarray}
e^{2\alpha(r_0)}=1-\frac{\varepsilon r_0 ^3}{3},
\end{eqnarray}
we obtain the interior metric
 \begin{eqnarray}  \label{int}
\begin{aligned}
g _{\mbox{int}} =&-\left(\frac{3}{2}
\sqrt{1-\frac{\varepsilon}{3} r_0 ^2}
-\frac{1}{2}\sqrt{1-\frac{\varepsilon}{3} r^2}\right)^2 dt ^2\\
&+ \left(1-\frac{\varepsilon}{3}r^2\right)^{-1} dr ^2
           +r ^2 \left(d \theta ^2 +\sin ^2 \theta d \psi ^2\right)
\end{aligned}
 \end{eqnarray}
for $r \leq r_0$. Now it is possible that there exist spherically
symmetric stars whose energy density $\varepsilon$ and radius $r_0$
satisfy
 \begin{eqnarray}
\sqrt{\frac{8}{3\varepsilon}} \leq  r_0 <\sqrt{\frac{3}{\varepsilon}}.
\label{collapse1}
 \end{eqnarray}
From (\ref{cein}), we know that the interior metric $g_{\mbox{int}}$
has singularity at \begin{eqnarray} \label{sing1} r_* =3 \sqrt{r_0 ^2 -\frac{8}{3
\varepsilon}}. \end{eqnarray}  where the pressure function $p(r)$ and the
scalar curvature $R$ blow up. When $r_0
>\sqrt{\frac{8}{3\varepsilon}}$, the singularity appears at $r_* \neq 0$ and
the origin $r=0$ is regular. However, for $r_0
=\sqrt{\frac{8}{3\varepsilon}}$, we have $r_* = 0$ and the
singularity appears at the origin.

In \cite{G}, Gu used the following step function as energy density to
study the gravitational collapse:
\[ \varepsilon(r)=\left\{
\begin{array}{ll}
\lambda \varepsilon & \ \mbox{for} \quad 0 \leq r < A\\
\mu \varepsilon &\ \mbox{for} \quad A \leq r < B\\
\varepsilon & \ \mbox{for} \quad B \leq r < r_0\\
0 &\ \mbox{for} \quad r \geq r_0
\end{array}\right .\]
where $0<A<B<r_0$, $\varepsilon$, $\lambda$, $\mu$ are three positive
constants and $\lambda>1$, $0<\mu<1$. The physical interpretation of the
above energy density is that the star has a considerable dense shell.

We shall see that, in certain circumstances, there exists a
singularity where the energy density is finite but the pressure
blows up.

Set $A=\sqrt[3]{\frac{1-\mu}{\lambda-\mu}}B$. Thus we have
\begin{equation}\label{modi}
\frac{1}{2}\int_0^A\lambda \varepsilon r^2 dr+\frac{1}{2}\int_A^B\mu
\varepsilon r^2 dr=\frac{1}{2}\int_0^B \varepsilon r^2
dr.\end{equation} Therefore, in the region $B<r<r_0$, the star has
mass function
\begin{eqnarray*}
\begin{aligned}
m(r)&=\frac{1}{2}\int_0 ^r \varepsilon(s) s^2 ds\\
&= \frac{1}{2}\int_0 ^A \lambda \varepsilon s^2 ds+\frac{1}{2}\int_A^B\mu
\varepsilon s^2 ds +
\frac{1}{2}\int_B ^r \varepsilon s^2 ds\\
&= \frac{1}{2}\left(\int_0^B+\int_B^r \right) \varepsilon s^2 ds\\
&=\frac{\varepsilon}{6}r^3
\end{aligned}
\end{eqnarray*}
and the total mass
$M=\frac{\varepsilon}{6}r_0^3$. By Birkhoff theorem \cite{ABS},
outside the star, the spacetime metric is the Schwarzschild solution
(\ref{ext}).

Similar to the constant energy density case \cite{OS}, plugging the
mass formula in the region $B<r<r_0$ back into the TOV (\ref{TOV})
with boundary condition $p(r_0)=0$, we see that the pressure has the
same expression (\ref{pre}). If
 \begin{eqnarray*}
r_* =3 \sqrt{r_0 ^2-\frac{8}{3\varepsilon}}>B,
 \end{eqnarray*}
there is also a spacetime metric
singularity where the energy density is finite but the pressure
and the scalar curvature blow up.

\subsection{Quadratically decaying energy density functions}

In \cite{MZ}, Misner and Zapolsky found an exact solution of the TOV
equation (\ref{TOV}) when the energy density function decays
quadratically. However, their solution does not satisfy the zero
pressure condition $p(r_0)=0$ on the boundary $r=r_0$ of the star.
Aided by information on the Riccati and Bernoulli equations, Zhong \cite{Z}
obtained an exact interior solution for the perfect fluid sphere. In
both of the above cases, the energy density is singular at the
origin $r=0$ and the physics for such a case is somewhat unclear. In
this subsection, by modifying the energy density function to be
finite near the origin, we propose a new model in which the energy
density is a decreasing function. We shall see that, inside the star, there exists
a real spacetime singularity and thus the origin is not naked.

Let $\varepsilon$ and $\lambda$ be two small positive constants such
that $0<\varepsilon<\frac{1}{2}$. Define the energy density function by
\[ \varepsilon(r)=\left\{
\begin{array}{ll}
\lambda & \ \mbox{for} \quad 0 < r \leq
\sqrt{\frac{3\varepsilon}{\lambda}}\\
\frac{\varepsilon}{r^2} &\ \mbox{for}
\quad \sqrt{\frac{3\varepsilon}{\lambda}} < r \leq r_0\\
0 &\ \mbox{for} \quad r > r_0
\end{array}\right .\]
Note that $\varepsilon(r)$ is a decreasing function and
$$\lim_{r\rightarrow \sqrt{\frac{3\varepsilon}{\lambda}}^-}
\varepsilon(r)
=\lambda>\frac{\lambda}{3}=\lim_{r\rightarrow
\sqrt{\frac{3\varepsilon}{\lambda}}^+}\varepsilon(r).$$
Thus the mass function is
\[ m(r)=\left\{
\begin{array}{ll}
\frac{\lambda}{6} r^3 & \ \mbox{for} \quad 0 < r \leq
\sqrt{\frac{3\varepsilon}{\lambda}}\\
\frac{\varepsilon }{2} r&\ \mbox{for} \quad
\sqrt{\frac{3\varepsilon}{\lambda}} < r \leq r_0\\
\frac{\varepsilon r_0}{2} &\ \mbox{for} \quad r > r_0.
\end{array}\right .\]

Similar analysis leads to the result that, outside the star (for $r > r_0$), the
spacetime metric is the Schwarzschild solution
\begin{equation}\label{Sch2}
\begin{aligned}
g_{\mbox{ext}}=&-\left(1-\frac{\varepsilon r_0}{r}\right)dt ^2 +
\left(1-\frac{\varepsilon r_0}{r}\right)^{-1} dr ^2\\
           &+r ^2 \left(d \theta ^2 +\sin ^2 \theta d \psi
          ^2\right)
\end{aligned}
\end{equation}
with the total mass
$\frac{\varepsilon r_0}{2}$.
In the region
$\sqrt{\frac{3\varepsilon}{\lambda}}<r<r_0$, the metric reads
\begin{eqnarray} \label{int3}
g_{\mbox{int}}&=
-e^{2\alpha(r)}dt^2+\frac{1}{1-\varepsilon}dr^2
+r^2(d\theta^2+\sin^2\theta
d\psi^2),
\end{eqnarray}
where $e^{\alpha(r)}$ can be described as follows. Let
\begin{eqnarray} q_+=\sqrt{1-\varepsilon}+\sqrt{1-2\varepsilon}, \quad
q_-=\sqrt{1-\varepsilon}-\sqrt{1-2\varepsilon}.
 \end{eqnarray}
Then
\begin{eqnarray*}
\begin{aligned}
e^{\alpha(r)}&=\frac{r^{1+\sqrt{\frac{1-2\varepsilon}{1-\varepsilon}}}
r_0 ^{-1+\sqrt{\frac{1-2\varepsilon}{1-\varepsilon}}}}{4\sqrt{1-2\varepsilon}}
\Big(q_+^2
r^{-2\sqrt{\frac{1-2\varepsilon}{1-\varepsilon}}}
-q_-^2
r_0^{-2\sqrt{\frac{1-2\varepsilon}{1-\varepsilon}}}\Big).
\end{aligned}
\end{eqnarray*}
Moreover, equation \eqref{TOV} can be integrated with boundary
condition $p(r_0)=0$ to yield the pressure
 \begin{eqnarray} \label{pre2}
p(r)=\frac{\varepsilon^2\Big(r^{-2\sqrt{\frac{1-2\varepsilon}{1-\varepsilon}}}
-r_0 ^{-2\sqrt{\frac{1-2\varepsilon}{1-\varepsilon}}}\Big)}
{r^2\Big(q_+^2
r^{-2\sqrt{\frac{1-2\varepsilon}{1-\varepsilon}}}
-q_-^2 r_0
^{-2\sqrt{\frac{1-2\varepsilon}{1-\varepsilon}}}\Big)},
 \end{eqnarray}
which is valid in the region
$\sqrt{\frac{3\varepsilon}{\lambda}}<r<r_0$ and is obviously
positive. The pressure $p(r)$ has finite limit as $r\rightarrow
\sqrt{\frac{3\varepsilon}{\lambda}}^+$.

The energy density $\varepsilon(r)$ is discontinuous at
$r=\sqrt{\frac{3\varepsilon}{\lambda}}$, but we require the pressure
be continuous at this point. By contracting the Einstein equation
(\ref{cein}), we obtain
 \begin{eqnarray}
\breve{P}_{-}:=\lim_{r\rightarrow
\sqrt{\frac{3\varepsilon}{\lambda}}^-}p(r)=\frac{2\lambda}{9}+\lim_{r\rightarrow
\sqrt{\frac{3\varepsilon}{\lambda}}^+}p(r).
 \end{eqnarray}
In the region $0<r<\sqrt{\frac{3\varepsilon}{\lambda}}$, the
pressure $p(r)$ satisfies the equation
\begin{equation}\label{TOV22}
\frac{dp}{(\lambda+p(r))(\lambda+3p(r))}=\frac{dr}{2(\lambda
r^2-3)}
\end{equation}
with boundary condition
 \begin{eqnarray*}
p\big|_{r=\sqrt{\frac{3\varepsilon}{\lambda}}}=
\frac{2\lambda}{9}+\frac{\varepsilon^2\Big[(\frac{3\varepsilon}
{\lambda})^{-\sqrt{\frac{1-2\varepsilon}{1-\varepsilon}}}
-r_0 ^{-2\sqrt{\frac{1-2\varepsilon}{1-\varepsilon}}}\Big]}
{\frac{3\varepsilon}{\lambda}\Big[q_+^2 (\frac{3\varepsilon}
{\lambda})^{-\sqrt{\frac{1-2\varepsilon}{1-\varepsilon}}}
-q_-^2 r_0 ^{-2\sqrt{\frac{1-2\varepsilon}{1-\varepsilon}}}\Big]}.
 \end{eqnarray*}

As in the case of constant energy density, one can obtain an explicit solution
of equation (\ref{TOV22}) with singularity at
\begin{eqnarray} \label{sing2}
r_\ast=\sqrt{\frac{3}{\lambda}-\frac{9(3-3\varepsilon)(\breve{P}_{-}
+\lambda)^2}{\lambda
(3\breve{P}_{-} +\lambda)^2}}.
\end{eqnarray}
Curvature tensors are finite but discontinuous at
$r=\sqrt{\frac{3\varepsilon}{\lambda}}$. So the point
$r=\sqrt{\frac{3\varepsilon}{\lambda}}$ is not a spacetime
singularity.

Note that the density function in reference \cite{Z} is of the form
$\rho=\frac{3c^2}{56\pi G r^2}$ (\cite[equation (4)]{Z}), where $c$
and $G$ denote the speed of light and Newton's constant
respectively. Thus, the coefficient of $\frac{1}{r^2}$ is fixed and
the only free parameter is the star radius $r_0$. In our context,
both $\varepsilon$ and $r_0$ are free parameters. If we set the
parameter $\varepsilon=\frac{3}{7}$, then the exterior Schwarzschild
solution (\ref{Sch2}) with the total mass $\frac{\varepsilon
r_0}{2}=\frac{3r_0}{14}$ can be glued to the interior
solution (\ref{int3}) at the boundary $r=r_0$ of the star, i.e.
 \begin{eqnarray*}
g=-\frac{4}{7}dt^2+\frac{7dr^2}{4}+r^2 \left(d
\theta ^2 +\sin ^2 \theta d \psi
           ^2\right).
 \end{eqnarray*}
This recovers (27) and (28), the particular case considered in \cite{Z}.

\section{Quantising interior solutions}\label{Interior}

In this section we quantise the metrics (\ref{int}) (see Section \ref{Decreasing})
following the approach to noncommutative general relativity developed
in \cite{CTZZ, WZZ1,WZZ2}. The purpose of our investigation is to examine
possible effects of space-time noncommutativity by considering simple models.

Note that the interior metrics (\ref{int}) and (\ref{int3}) for
spherically symmetric stars can both be written as
\begin{eqnarray}
g_{\mbox{int}}=-a^2(r)dt^2 +b^2(r)dr^2 +r^2(d\theta^2+\sin^2\theta
d\psi^2), \quad r \leq r_0,  \label{NCg}
\end{eqnarray}
for appropriate functions $a(r)$ and $b(r)$ where
\begin{eqnarray}
a(r_*)=0, \quad a'(r_*) \neq 0, \quad a''(r_*) \neq 0, \quad b(r_*)
\neq 0, \quad b'(r_*) \neq 0 \label{NCab}
\end{eqnarray}
for some $r_* <r_0$.

To carry out the deformation quantisation (in the sense of
\cite{CTZZ, WZZ1,WZZ2}) of the the metric \eqref{NCg}, we need first
to specify a Moyal algebra. Denote $x^0=t$, $x^1=r$, $x^2=\theta$
and $x^3=\phi$. We deform the algebra of functions in these
variables by imposing the Moyal product
\begin{eqnarray*}
(f\ast g)(x) = f(x) \exp^{\frac{\bar h}{2}\theta^{\mu \nu}
\stackrel{\longleftarrow}{\partial_\mu}\
\stackrel{\longrightarrow}{\partial_\nu}}g(x)
\end{eqnarray*}
with the following anti-symmetric matrix
\begin{eqnarray}\label{ansatz}
\left(\theta^{\mu \nu}\right)_{\mu,
\nu=0}^3=\left(\begin{array}{cccc}
   0&  0&  0&  0\\
   0&  0&  0&  0\\
   0&  0&  0&  1\\
   0&  0&  -1&  0
\end{array}\right),\label{asy-matrix}
\end{eqnarray}
where $\bar h$ is the deformation parameter, which may be regarded as
related to the Planck constant.

Some comments are in order. Noncommutativity of time coordinate
results in violation of unitarity and causality for quantum field
theories defined on flat spacetimes as well as spacetimes with
compact spatial submanifolds \cite{CDPT} (also see \cite{GM};
for a discussion in the context of string theory, see \cite{SST}).
In order to retain basic principles of
quantum physics such as unitarity and causality, we have to keep the
time coordinate commutative, and the Ansatz \eqref{ansatz} enforces
this.  Note also that the Ansatz \eqref{ansatz} leads to the
simplest possible model of space-time noncommutativity.

Following \cite{CTZZ}, we choose the following embedding
\begin{eqnarray}\label{nc-embed}
\begin{aligned}
& X^1=a(r) \sin t, \;\;X^2=a(r) \cos t,\;\;X^3=f(r), \\
& X^4=r \sin\theta \cos\phi,\;\;X^5
=r \sin\theta \sin\phi,\;\;X^6=r \cos\theta,
\end{aligned}
\end{eqnarray}
where $f(r)$ is related to the functions $a(r)$ and $b(r)$ through
the differential equation
\[(f')^2 +1=(a')^2 +b^2.\]

Here the rationale is much the same as in the classical theory of
surfaces, which can be treated as two-dimensional geometries
embedded in the three-dimensional Euclidean space. The surfaces
themselves are determined by the embeddings, and their differential
geometry can be studied by analyzing the embeddings in a very
concrete manner using elementary techniques (see, e.g., \cite{doC}
for an introduction to the theory of surfaces).

The embedding \eqref{nc-embed} determines the noncommutative
geometry of the space-time, which we now analyze. The quantum
deformation of the metric (\ref{NCg}) is defined by
\begin{eqnarray}\label{deformed-metric}
\begin{aligned}
\g_{\mu \nu} =& -\partial_\mu X^1\ast \partial_\nu X^1
-\partial_\mu X^2\ast \partial_\nu X^2
+\partial_\mu X^3\ast \partial_\nu X^3  \\
&+ \partial_\mu X^4\ast \partial_\nu X^4
+\partial_\mu X^5\ast \partial_\nu X^5
+\partial_\mu X^6\ast \partial_\nu X^6.
\end{aligned}
\end{eqnarray}
Note that in the classical commutative limit with $\bar h=0$,
\eqref{deformed-metric} reduces to
$g_{\mu\nu}=-\sum_{i=1}^2\partial_\mu X^i \partial_\nu X^i
+\sum_{j=3}^6\partial_\mu X^j \partial_\nu X^j$, which indeed
recovers the metric (\ref{NCg}) as one can verify.

Lengthy computations yield the following result for the
noncommutative metric \eqref{deformed-metric}:
\begin{eqnarray}\label{deformed-NCg}
\begin{aligned}
\g_{0 0} =&-a^2(r),\\
\g_{0 1} =&\g_{1 0}=\g_{0 2} =\g_{2 0}=\g_{0 3} =\g_{3 0}=0, \\
\g_{1 1} =&b^2(r)+(\sin^2\theta -\cos^2 \theta)\sinh^2\barh,\\
\g_{1 2} =&\g_{2 1} = 2r\sin\theta\cos\theta\sinh^2\barh,\\
\g_{1 3} =&-\g_{3 1} = -2r\sin\theta\cos\theta\sinh\barh\cosh\barh,\\
\g_{2 2} =&r^2
    \left[1-\left(\sin^2\theta-\cos^2\theta\right)\sinh^2\barh\right],\\
\g_{2 3} = &-\g_{3 2}
=r^2\left(\sin^2\theta-\cos^2\theta\right)\sinh\barh\cosh\barh,\\
\g_{3 3} =&r^2
    \left[\sin^2\theta+\left(\sin^2\theta-\cos^2\theta\right)\sinh^2\barh\right].
\end{aligned}
\end{eqnarray}

Given the general theory of \cite{CTZZ}, the computation of the
connection and Riemaniann curvature of the noncommutative metric
(\ref{deformed-NCg}) is in principle straightforward though very
laborious. We shall not spell out the complete results here.
However, we shall need the noncommutative scalar curvature, which
can be expressed in the form
 \begin{eqnarray}\label{As}
{\bf R}=\frac{A_0}{2 r^2A^3}-\frac{a'(r) A_1}{r a(r) A^2}-\frac{a''(r) A_2}{a(r) A},
 \end{eqnarray}
where $A$, $A_0$, $A_1$ and $A_2$ are given in the Appendix.

For the sake of being concrete, we shall only discuss the
noncommutative singularity for quantum analogue of the
metric (\ref{int}). As we have seen in Section \ref{Decreasing},
singularity occurs essentially in the region with
constant energy density. Similarly in the present setting,  if the
energy density $\varepsilon$ and radius $r_0$ of a spherically
symmetric satisfy (\ref{collapse1}), noncommutative singularity also occurs
for the metric (\ref{NCg}) at
 \begin{eqnarray*}
r_* =3 \sqrt{r_0 ^2 -\frac{8}{3 \varepsilon}}.
 \end{eqnarray*}
Indeed, in this case,
\begin{eqnarray}\label{solution}
\begin{aligned}
a(r)&=\frac{3}{2}
\sqrt{1-\frac{\varepsilon}{3} r_0 ^2}
-\frac{1}{2}\sqrt{1-\frac{\varepsilon}{3} r^2}, \\
b(r)&= \left(1-\frac{\varepsilon}{3}r^2\right)^{-\frac12},
\end{aligned}
\end{eqnarray}
we obtain
\begin{eqnarray}\label{Bs}
{\bf R} = B_1 - \frac{B_2}{B},
\end{eqnarray}
where $B$, $B_1$ and $B_2$ are given in the Appendix.

Note that $B|_{r=r_*}=0$. Inspecting the formulae for $B_1|_{r=r_*}$ and $B_2|_{r=r_*}$ in the Appendix,
we see that $B_1|_{r=r_*}$ is finite and $B_2|_{r=r_*} \ne 0$. Thus
we have
\begin{eqnarray}\label{Brstars}
{\bf R}(r_*) = \infty,
\end{eqnarray}
and hence $r=r_*$ is a singularity
in the noncommutative spacetime.

\begin{rmk}
If $r_0^2 \varepsilon>\frac{8}{3}$, then $r=0$ is a regular point of
the quantum deformed metric for (\ref{int}). In this case, the
noncommutative scalar curvature \begin{eqnarray*}  {\bf R} = -\frac{\varepsilon
\left(\cosh 2 \bar h-2 \sqrt{9-3 r_0^2 \varepsilon }+3\right)
   \mbox{sech} ^4 \bar h}{\sqrt{9-3 r_0^2 \varepsilon }-1}
\end{eqnarray*}
is finite. When $r_0 =\sqrt{\frac{8}{3\varepsilon}}$,
the origin $r=0$ is singular.
\end{rmk}

\section{Noncommutative gravitational collapse}\label{Collapse}

In this section, we quantise the dust solutions \cite{OS} and study
noncommutative gravitational collapse. While our method for quantisation
is much the same as in Section \ref{Interior}, a new feature
is that the quantised dust solutions have an explicit time dependence and
their time evolutions are thus clear.

By replacing $\tau $ by $t$, and $R$ by $r$,
the Tolman metric studied in \cite{OS} can be written as
\begin{eqnarray}
ds^2=-dt^2+(1-c t)^{4/3} \big[dr^2+r^2 (d\theta ^2+ \sin ^2\theta d\phi ^2)\big] \label{Oppenhm-S}
\end{eqnarray}
with $c=3 r_0 ^\frac{1}{2} R _b ^{-\frac{3}{2}}$, where $r_0$ is the gravitational radius and $R_b$
is the radius of the star or some micro object. This spacetime can be embedded into a 5-dimensional
flat Minkowski spacetime via
\begin{eqnarray}\label{five-X}
\begin{aligned}
X^1=&\frac{9 (1-c t)^{4/3}}{32 c^2}+\Big(\frac{r^2}{4}+1\Big) (1-c t)^{2/3},\\
X^2=&\frac{9 (1-c t)^{4/3}}{32 c^2}+\Big(\frac{r^2}{4}-1\Big) (1-c t)^{2/3},\\
X^3=&(1-c t)^{2/3} r \cos \phi  \sin \theta ,\\
X^4=&(1-c t)^{2/3} r \sin \theta  \sin \phi ,\\
X^5=&(1-c t)^{2/3} r \cos \theta .
\end{aligned}
\end{eqnarray}

As in Section \ref{Interior}, we deform the algebra of functions in the variables
$r, t, \pi$ and $\theta$ into a Moyal algebra ${\mathcal A}$ defined by the anti-symmetric matrix
(\ref{ansatz}). Now we consider the noncommutative geometry embedded in
${\mathcal A}^5$ by \eqref{five-X}. The noncommutative metric of the embedded noncommutative geometry
(defined in the standard way \cite{CTZZ})  yields a
quantum deformation of the metric (\ref{Oppenhm-S}):
\begin{eqnarray}
\begin{aligned}
\g_{\mu \nu} =& -\partial_\mu X^1\ast \partial_\nu X^1
+\partial_\mu X^2\ast \partial_\nu X^2
+\partial_\mu X^3\ast \partial_\nu X^3  \\
&+ \partial_\mu X^4\ast \partial_\nu X^4
+\partial_\mu X^5\ast \partial_\nu X^5,
\end{aligned}
\end{eqnarray}
which can be computed explicitly. We have
\begin{eqnarray*}
\begin{aligned}
\g_{11}=&-\frac{4 c^2 r^2 \cos  2 \theta  \sinh ^2\bar h}{9 (1-c t)^{2/3}}-1,\\
\g_{12}=&\g_{21}=\frac{2}{3} c r (1-c t)^{1/3}\cos  2 \theta  \sinh ^2\bar h,\\
\g_{13}=&\g_{31}=-\frac{4}{3} c r^2 (1-c t)^{1/3} \cos  \theta  \sin  \theta  \sinh ^2\bar h,\\
\g_{14}=&-\g_{41}=\frac{1}{3} c r^2 (1-c t)^{1/3} \sin  2 \theta  \sinh  2 \bar h,
\end{aligned}
\end{eqnarray*}
\begin{eqnarray*}
\begin{aligned}
\g_{22}=&(1-c t)^{4/3} \left(1-\cos  2 \theta  \sinh ^2\bar h\right),\\
\g_{23}=&\g_{32}=r (1-c t)^{4/3} \sin  2 \theta  \sinh ^2\bar h,\\
\g_{24}=&-\g_{42}=-2 r (1-c t)^{4/3} \cos  \theta  \cosh \bar h \sin  \theta  \sinh \bar h,\\
\g_{33}=&r^2 (1-c t)^{4/3} \left(\cos  2 \theta  \sinh ^2\bar h+1\right),\\
\g_{34}=&-\g_{43}=-\frac{1}{2} r^2 (1-c t)^{4/3} \cos  2 \theta  \sinh  2 \bar h,\\
\g_{44}=&-\frac{1}{2} r^2 (1-c t)^{4/3} (\cos  2 \theta  \cosh  2 \bar h-1).
\end{aligned}
\end{eqnarray*}
The noncommutative scalar curvature is given by
\begin{eqnarray}\label{Cs}
{\bf R}=\frac{4 c^2 \cosh ^2\bar h}{(1-c t)^{4/3}}\frac{C_1}{C^3}
\end{eqnarray}
where $C$ and $C_1$ are given in the Appendix.

We shall consider aspects of the noncommutative spacetime
by examining the behaviour of the scalar curvature as time increases by
following an approach adopted in \cite{W} in the classical context.
When time approaches values where ${\bf R}\to \infty$,
the radius of the stellar object reduces to zero,
and this is an indication of gravitational collapse \cite{W}.
Obviously this only provides a snapshot, nevertheless, it enables us to gain
some understanding of gravitational collapse in the non-commutative setting.
A full treatment of the time evolution of stellar objects ending
at gravitational collapses in noncommutative geometry will be
given in a future publication.

Let us regard $\bar h$ as a real number and make the (physically realistic)
assumption that $\bar h$ is positive but close to zero.
Now if $t$ is significantly smaller than $\frac{1}{c}$ compared to $\bar h$, that is,
$\frac{1}{c}-t\gg\bar h$, both the noncommutative metric and
noncommutative scalar curvature $\bf R$ are finite,
and there is non-singularity in the noncommutative spacetime.
Thus the stellar object described by the noncommutative geometry behaves
much the same as the corresponding classical object.

When $t=t_*:=\frac{1}{c}$, we have $\bf R | _ {t_{*} =\frac{1}{c}}=\infty$
and the radius of the stellar object reduces to zero. This is the time when
gravitational collapse happens in the usual classical setting.

However, in the noncommutative case,
singularities of the scalar curvature already appear
before $t_*$. Indeed, when time reaches
\[
\begin{aligned}
t(r, \theta)&=\frac{1}{c} -
\frac{\sqrt{8}}{27} c^2 r^3
(2 \cos  2 \theta +\cosh  2 \bar h+3)^{3/2} \frac{\sinh^3\bar h}{\cosh^6\bar h}\\
&\cong \frac{1}{c} -
\frac{8}{27} c^2 r^3
(\cos  2 \theta + 2)^{3/2} \bar h^3.
\end{aligned}
\]
$C$ vanishes and $C_1/(1-c t)^{4/3}$ is finite of order $0$ in $\bar h$.
Thus the scalar curvature tends to infinity for all $t(r, \theta)$
and the noncommutative spacetime becomes singular.

This indicates that in the noncommutative setting,
gravitational collapse happens within a certain range of time
because of the quantum effects captured by the non-commutativity of
spacetime.  This is fully consistent with usual
expectations of quantum mechanics. However,
effect of non-commutativity only starts to appear at third order
of $\bar h$.

\begin{appendix}
\section{Some formulae used in the main text}\label{Appendix}
This appendix spells out some lengthy formulae which have been used in the main body of the paper.
The quantities $A$, $A_0$, $A_1$ and $A_2$ used in \eqref{As} are given by
\begin{eqnarray*}
\begin{aligned}
A=&2 \Big(\cos 2 \theta \sinh ^2 \bar h-1\Big) b(r)^2
   -\Big(2 \cos 2 \theta +\cosh 2 \bar h +3\Big) \sinh ^2 \bar h,\\
A_0=&-8 \Big(3 \cos 2 \theta \sinh ^2 2 \bar h+2\cosh 2 \bar h+2\cosh 4 \bar h\Big) b(r)^6\\
    &+2 \Big[\big(-8 \cos 4 \theta \sinh ^4 \bar h+92 \cosh 2 \bar h-19 (\cosh 4 \bar h+3)\big) \cosh ^2 \bar h\\
    &+4 \cos 2 \theta (7-4 \cosh 2 \bar h) \sinh ^2 2 \bar h \Big] b(r)^4\\
    &+2 r \Big[(8 \cos 4 \theta \sinh ^4 \bar h-28 \cosh 2 \bar h+\cosh 4 \bar h-5) \cosh ^2 \bar h\\
   &+2 \cos 2 \theta (3 \cosh 2 \bar h-1) \sinh ^2 2 \bar h \Big] b'(r) b(r)^3\\
   &-4 \Big[4 \big(\cos 2 \theta (3 \cosh 2 \bar h-23)-2 \cos 4 \theta \big) \cosh ^2 \bar h \sinh ^4 \bar h\\
   &+(-14 \cosh 2 \bar h+\cosh 4 \bar h+9) \sinh ^2 2 \bar h \Big] b(r)^2\\
   &+r \Big[\cos 2 \theta (3 \cosh 4 \bar h+13) \sinh ^2 2 \bar h\\
   &-16 (\cos 4 \theta-\cosh 2 \bar h+2) \cosh ^2 \bar h \sinh ^4 \bar h \Big] b'(r) b(r)\\
   &+16\Big[-7 \cos 2 \theta-\cos 4 \theta+(3 \cos 2 \theta+4)\cosh 2 \bar h-5\Big] \cosh ^2 \bar h \sinh ^4 \bar h,\\
A_1=&8 \cos 2 \theta \cosh ^2 \bar h \sinh ^4 \bar h+\Big(3 \cosh 2 \bar h+1\Big)\sinh ^2 2 \bar h\\
    &+2 \Big[(6 \cosh 2 \bar h+2) \cosh ^2 \bar h+\cos 2 \theta \sinh ^2 2 \bar h\Big] b(r)^2\\
    &-8 r \Big(\cos 2 \theta \sinh ^2 \bar h-1\Big)^2 b(r) b'(r),\\
A_2=&4 \Big(\cos 2 \theta \sinh ^2 \bar h-1\Big).
\end{aligned}
\end{eqnarray*}

The following quantities appeared in equation \eqref{Bs}:
\begin{eqnarray*}
\begin{aligned}
B=&-3+\varepsilon  r^2+3 \sqrt{(3-r^2 \varepsilon)(3-r_0^2 \varepsilon)},\\
B_1=&-\frac{\varepsilon}{2 \left(\Big[2 r^2 \varepsilon \cos  2 \theta
+\left(r^2 \varepsilon -3\right) (\cosh  2 \bar h +3)\Big] \sinh ^2 \bar h -6\right)^3}\\
& \times \Big\{  \Big[-16 r^4 \varepsilon ^2 \cos  4 \theta  \sinh ^4 \bar h
-2 r^2 \varepsilon  \left(23 \varepsilon  r^2+9\right)\\
&+\left(\varepsilon r^2 \left(68 \varepsilon
   r^2+21\right) +405\right) \cosh  2 \bar h
   +2 \left(\varepsilon  r^2 \left(9-13 r^2 \varepsilon \right)+81\right) \cosh  4 \bar h\\
&+\left(r^2 \varepsilon -3\right) \left(4 r^2
   \varepsilon -9\right) \cosh 6 \bar h +270\Big] \cosh ^2 \bar h \\
   &+2 r^2 \varepsilon  \cos  2 \theta  (3 \cosh  2 \bar h -7)
   \left(\left(r^2 \varepsilon -3\right) \cosh 2 \bar h -3-\varepsilon  r^2\right)
   \sinh ^2 2 \bar h \Big\},\\
B_2=&-\frac{1}{\left(\Big[2 r^2 \varepsilon  \cos  2 \theta
+\left(r^2 \varepsilon -3\right) (\cosh  2 \bar h +3)\Big] \sinh
   ^2 \bar h -6\right)^2}\\
& \times \Big\{\varepsilon  \left(r^2 \varepsilon -3\right)
\cosh ^2 \bar h  \Big[-8 r^2 \varepsilon  \cos  2 \theta  \sinh ^4 \bar h -r^2 \varepsilon\\
&+4 \left(\varepsilon  r^2+9\right) \cosh 2 \bar h +\left(9-3 r^2 \varepsilon \right)
\cosh  4 \bar h +27\Big]\Big\}.
\end{aligned}
\end{eqnarray*}

Evaluating $B_1$ and $B_2$ at $r=r_*=3 \sqrt{r_0 ^2 -\frac{8}{3 \varepsilon}}$, we obtain
\begin{eqnarray*}
\begin{aligned}
B_1|_{r=r_*}=&-\frac{\varepsilon}{2 \left(\Big[6 \left(3 r_0^2
\varepsilon -8\right) \cos 2 \theta+9 \left(r_0^2
   \varepsilon -3\right) (\cosh 2 \bar h+3)\Big] \sinh ^2\bar h-6\right)^3}\\
   &\times \Big\{9 \cosh ^2\bar h\Big[\left(612 \varepsilon ^2 r_0^4-3243
   \varepsilon  r_0^2+4341\right) \cosh 2 \bar h\\
   &-2 \left(3
   \varepsilon  \left(39 r_0^2 \varepsilon -211\right) r_0^2+847\right) \cosh 4 \bar h\\
   &+3 \left(r_0^2 \varepsilon -3\right)
   \left(12 r_0^2 \varepsilon -35\right) \cosh 6 \bar h\\
   &-2 \left(8 \left(8-3 r_0^2 \varepsilon \right)^2 \cos 4 \theta \sinh
   ^4\bar h+3 r_0^2 \varepsilon  \left(69 r_0^2 \varepsilon -365\right)+1433\right)\Big] \\
   &+18 \sinh ^2 2 \bar h\Big[ \left(3 r_0^2
   \varepsilon -8\right) \cos 2 \theta \left(3 \cosh 2 \bar h-7\right)\\
   &\times\left(-3 \varepsilon  r_0^2+3 \left(r_0^2 \varepsilon -3\right) \cosh
   2 \bar h+7\right)\Big]\Big\},
\end{aligned}
\end{eqnarray*}
\begin{eqnarray*}
\begin{aligned}
B_2|_{r=r_*}=&-\frac{3 \varepsilon  \left(r_0^2 \varepsilon
-3\right) \cosh ^2\bar h} {\left(\Big[2 \left(3 r_0^2 \varepsilon
   -8\right) \cos 2 \theta+3 \left(r_0^2 \varepsilon -3\right) (\cosh 2 \bar h+3)\Big]
   \sinh ^2\bar h-2\right)^2}\\
   &\times\Big[8 \left(8-3r_0^2 \varepsilon \right)
   \cos 2 \theta \sinh ^4\bar h-3 r_0^2 \varepsilon \\
   &+4\left(3 r_0^2 \varepsilon -5\right)
   \cosh 2 \bar h-9 \left(r_0^2 \varepsilon
   -3\right) \cosh 4 \bar h+17 \Big].
\end{aligned}
\end{eqnarray*}
These formulae are used in the derivation of \eqref{Brstars}.

The quantities $C$ and $C_1$ in \eqref{Cs} in the main body of the paper are given by the following formulae:

\begin{eqnarray*}
\begin{aligned}
C=&9(1-c t)^{2/3} \cosh ^4\bar h-2 c^2 r^2 (2 \cos  2 \theta +\cosh  2 \bar h+3) \sinh ^2\bar h,
\end{aligned}
\end{eqnarray*}
\begin{eqnarray*}
\begin{aligned}
C_1 =&-243 (1-c t)^{4/3} \cosh ^8\bar h+486 (1-c t)^{4/3} \cosh ^6\bar h\\
   &-18 c^2 r^2 (1-c t)^{2/3} (2 \cos  2 \theta -3 \cosh  2 \bar h-1) \sinh ^2\bar h \cosh^4\bar h\\
   &-9 c^2 r^2 (1-c t)^{2/3} \Big(52 \cosh  2 \bar h+3 \cosh  4 \bar h\\
   &+\cos  2 \theta  (28 \cosh  2 \bar h+\cosh  4 \bar h-13)+9\Big) \sinh ^2\bar h \cosh ^2\bar h\\
   &+4 c^4 r^4  \sinh^4\bar h \Big(4 \cos  2 \theta  (\cosh  2 \bar h+15) \sinh ^2\bar h\\
   &+2 \cos  4 \theta  (\cosh  2 \bar h-3)+38 \cosh  2 \bar h+3 \cosh  4 \bar h-13\Big).
\end{aligned}
\end{eqnarray*}

\end{appendix}

\medskip

\noindent {\bf Acknowledgement.} N. Xie wishes to thank Profs. C.H.
Gu and H.S. Hu for their encouragement. X. Zhang wishes to thank the
School of Mathematics and Statistics, University of Sydney for the
hospitality during his visits when part of this work was carried
out. Partial financial support from the Australian Research Council,
National Science Foundation of China (grants 10421001, 10725105,
10731080, 10801036), NKBRPC (2006CB805905) and the Chinese Academy
of Sciences is gratefully acknowledged.

\end{document}